# MERGING SUPPLY CHAIN AND BLOCKCHAIN TECHNOLOGIES

ELJAZZAR, M. M AND AMR, M. A

KASSEM, S. S. AND EZZAT, M.

**ABSTRACT:** Technology has been playing a major role in our lives. One definition for technology is "all the knowledge, products, processes, tools, methods and systems employed in the creation of goods or in providing services". This makes technological innovations raise the competitiveness between organizations that depend on supply chain and logistics in the global market. With increasing competitiveness, new challenges arise due to lack of information and assets tractability. This paper introduces three scenarios for solving these challenges using the Blockchain technology. In this work, Blockchain technology targets two main issues within the supply chain, namely, data transparency and resource sharing. These issues are reflected into the organization's strategies and plans.

*KEYWORDS:* **SUPPLY CHAIN, BLOCKCHAIN**

## 1- INTRODUCTION

Since the rise of the 20$^{th}$ century, the dependency of businesses on supply chains has increased. Supply chains have become a vital business process that can encourage the competitiveness of a business in the market. It has been recently emphasized that the success of a company is based on the interaction between flows of information, materials, workers and capital equipment. Logistics have been playing a major role in achieving an effective value proposition to the customers as defined by Cavinato Kshetri, N. (2018). The author clearly mentioned that retail business logistics are essential for creating business value as he defined the logistics as "The integrated management of purchasing, transportation and storage on functional basis".

Supply chain management was used for decades to deliver goods. Later, the supply chain management concept changed from focusing just on logistics to become mandatory to track and manage other flows within the supply chain. Such flows include, flow of money and flow of information. There are billions of products being manufactured every day through complex systems of supply chains that extend to all parts of the world.. However, there are several questions regarding these products. For example, how, when and where these products were originated, manufactured, and used through their life cycle. Even before reaching the end consumer, goods pass through a wide network of retailers, distributors, transporters, inventories, and suppliers that participate in every process in the supply chain (i.e. design, production, delivery, sales, etc...).

Most of these processes need quick actions. Unfortunately, little information is available during the supply chain life cycle. This is one of major challenges in the supply chain domain. Blockchain – a kind of distributed ledger technology—has been described as the next industry revolution. Blockchain is a data structure format that makes it possible to create a stack of existence with a specific time stamp for each digital asset. This could be helpful to sign each transaction in a distributed digital ledger Lummus, R. R (2001). Due to the above challenges, the main purpose of this paper is to introduce the merge between Blockchain and supply chain. The rest of the paper is organized as follows: In section 2, previous work considering adopting Blockchain in the supply chain domain is presented. In Section 3, investigates different applications of Blockchain. In section 4, a short description of implementation is given. Finally Section 5 provides conclusion of the paper.

## 2- LITERATURE REVIEW

Supply chains are getting increasingly more complex, more extended, and more global, due to the huge advance in technology, and the more complex production systems. With the rapid growth of internet of things, many researchers consider the application of relevant technologies for systems traceability in supply chains. Folinas et al pointed out that the efficiency of a traceability system depends on the ability to track and trace each digital unit (individual product and logistics units), in a way that enables real time monitoring from primary production until final delivery to the consumer In Tjandra, R. A., & Tan, K. H. (2002, September). While Mattoli et al. developed a Flexible Tag Data-logger (FTD) based on RFID. Which is attached to the bottles for collecting some metrological data, (light, humidity, and temperature). in order to trace the bottles to a supermarket. The historical data stored in the FTD system that can be read by smart phone with integrated infrared port to evaluate the safety status of wine bottles in Folinas, D., Manikas, I., & Manos, B. (2006).

The digital supply chain is described by the exchange of information between suppliers (financial, production, design, research, and/or competition) to enhance communication between stakeholders in the chain. As every piece of information stored affects the forecasting in supply chain, several models are developed to collect these features (gathered information), then optimize the forecasting Mattoli, V (2009). This in turn, reflects the accuracy of forecasting models to supply chain demand.

With the age of Internet of Things, the flow of information sharing and processing is not confined to the business process level only, but also it includes a huge amount of data from low resourced users (devices and sensors and from social media applications). It becomes essential to detect casual events when forecasting, as mentioned, in a previous work considering the weekends, public holidays, sudden events which are reflected in the decision-making and actions in 6. Eljazzar, M. M., & Hemayed, E. E. (2016, December). Integrated supply chain information models are an essential feature at the current time. The role of information integration and service automation has been identified as an important business driver.

All the previous work is based on the idea of using a data centralized system. Which was, until recently, the main conceivable approach to achieve information transparency along supply chains. With the raise of Blockchain, a whole new approach has drawn attention from researchers in many different domains in public services, Internet of Things (IoT), smart cities, , reputation systems and

security services Elgazzar, M. M., & Hemayed, E. E. (2016, December), Azaria, A., Ekblaw, A (2016, September).

Supply chain is not far away from Blockchain. Zyskind, G., & Nathan, O. (2015, May). discussed the benefit of merging Blockchain technology in manufacturing supply chain. They listed the main characteristics of the Blockchain software architecture to enhance trust through transparency and traceability in the supply chain network. In addition to the transparency within any transaction of data, goods, and financial resources, Blockchain could offer an innovative network for new decentralized and transparent transaction mechanism in industries and business. The authors also use an example to demonstrate how Blockchain can be used in global supply chain networks; it acts as a spark to encourage researchers to introduce their work in this domain. 11. McConaghy, et all (2016). Introduced the BigchainDB, which acts as a distributed database (DB) with Blockchain features. Therefore, it has characteristics of distributed databases: linear scaling in throughput and capacity optimized querying. Moreover, the main Blockchain feature is the decentralized, immutability for creation & movement of digital assets.

According to the discussion above, internet of things has been widely used for supply chain
traceability systems. However, most of them are centralized systems that suffer from several issues,
which could be overcome using the Blockchain technology. Therefore, the contribution of (Nir Kshetri, 2018) discussed the roles of Blockchain in meeting the objectives of the supply chain
management through exploring different case studies and initiatives In the next section, the benefits
of merging Blockchain with supply chain is presented. These benefits include: cost-effectiveness of
services and value-creating activities that are advantageous to many actors in the ecosystem, for example, firms and their suppliers, employees and customers. The benefit of Blockchain in the supply chain could be seen in two perspectives, namely, data transparency and resource sharing.

### 3- DATA TRANSPARENCY

The main advantage and impact of the Blockchain is the trust of transactions and the ability to share assets. Every transaction is trusted and secured due to the presence of ledgers. A ledger is is a consensus of replicated, shared, and synchronized digital data geographically spread across multiple sites, countries, or institutions. There is no central administrator or centralized data storage. Accordingly, Blockchain users are provided with the ability to share assets without concerns regarding how to protect them. In this section, a short a review for two scenarios on how information flows through the supply chain could be empowered using Blockchain.

3.1 Enhancing information:

Managing the information in the supply chain is very critical in the supply chain processes since it is the main component that affects supply chain planning. Supply chain planning depends on demand forecasting, transportation, inventory management… etc. Due to the data transparency of the distributed ledger (Blockchain); the trust is achieved between users. This is reflected as the flow of information between users and supply chain parties for information sharing as transactions or digital assets in the Blockchain network. This step leads to enhancing both the level of information and coordination among the supply chain, leading to a more win-win situations towards increasing the profitability among the supply chain. This is possible since data availability with time stamps allow for better decision making.

One of the pioneers in this step is Maersk, one of the biggest containers' carrier in the world. They apply a solution that depends on the Blockchain in the international market logistics to track their containers, and track the required paperwork needed at the customs and ports. The transportation of this paperwork may cost as much as the transportation of the shipment. In addition, the information can be exposed to fraud when it is delivered physically. At the end of 2017, they were able to get the customs to try the solution, upload this amount of paperwork online, and digitally copy it.

3.2 Demand forecasting

The quality of demand forecasting depends on the amount and the quality of shared information among the supply chain parties. Hence, the more data in a supply chain, the more accurate forecasts are made. One of the main characteristics of Blockchain technology is that it is an append only system, where, every single transaction in linked in the Blockchain network with a time stamp. The Blockchain can enhance and increase the quality of these pieces of information to increase the accuracy of the Sales and Operation planning (S&OP).

In 2016, it was reported that Walmart was testing a service to monitor pork in China. The solution involved moving pork products from Chinese farms to Chinese stores. As of February 2017, it had completed the trials. Walmart was reported to be confident that a finished version could be ready "within a few years". Blockchain enabled the digitally track individual pork products in a few minutes compared to many days. Details about the farm, factory, batch number, storage temperature and shipping can be viewed on blockchain as a distributed ledger. These details helped to plan and forecast for the S&OP.

## 4- RESOURCE SHARING

One of the main advantages of Blockchain technology is the data transparency. This feature could help the decision maker to know which digital assets in the supply chain are fully utilized. This assists in planning for better utilization through sharing resources.

4.1 Sharing resources in Supply Chain networks

As mentioned before, one of the main edges of the Blockchain is that it enables the different parties to share their digital assets. Hence, it can provide the normal users to use their houses resources and empty small spaces as warehouses as well as trucks, vehicles and all their transportation assets. In addition, this scenario is not going to affect the normal flow of logistics but it will also help the different parties to enhance their reverse logistics plans and become involved in the network. In another scenario, a factory could share their production lines based on weekly action plans. The result is reflected on resource utilization and net profit, as shown in figure 1.

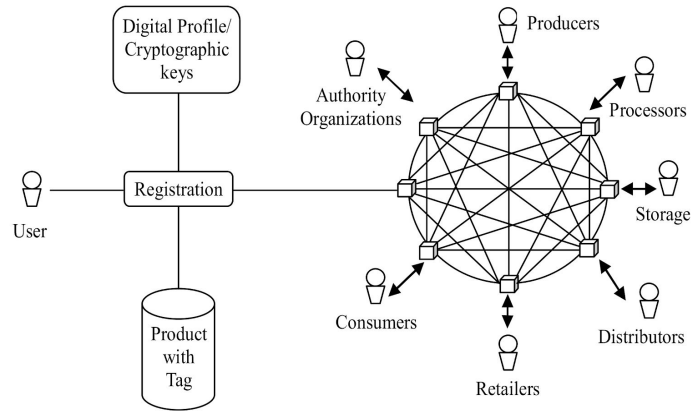

Figure 1. Interaction between vast users in the Blockchain network.

4.2 Flow of money (Barter)

Cryptocurrencies – especially Bitcoin - began to be used among many countries as a legal currency. Every day witness a birth of new crypto currency with different values. These currencies are used in trading exchanges. when one mention the flow of money in the supply chain; one could imagine that we will make another cryptocurrency for trading with new values. but in reality, what recommended from other works is the extinction of the currencies as people will later can exchange goods against goods. When using the Blockchain technology to connect more than two parties together so that they can find their mutual benefits leading to an era where the currency or money is not important as the true value that someone desires from owning goods.

At the end, each one of these scenarios can be enhanced to achieve a better value if we merge it with other technologies that would result as a disruptive solution to many supply chains.

## 5- CONCLUSIONS

Supply chain processes could be easily transformed to meet the Blockchain architecture. Blockchain facilitates valid and effective measurement of outcomes and performance of key supply chain processes through data transparency and information flow. Blockchain is an emerging field, and the merge between Blockchain and supply chain is promising. This is due to the fact that Blockchain has been recognized as promising software architecture in several domains other than supply chain. This paper proposed a 3 scenarios to merge Blockchain and supply chain in order to avoid some of the challenges faced by supply chains. The scenarios are based on Data Transparency and Resource sharing. The paper concludes that the proposed scenarios of merging Blockchain and supply chain technologies will enhance supply chains performance.